\documentclass[12pt,prd,aps,tightenlines]{revtex4-1}

\usepackage[dvips]{epsfig}
\usepackage{amssymb}
\usepackage{amsbsy}
\usepackage{chngcntr}
\usepackage{amsmath}
\usepackage{amsfonts}
\usepackage{verbatim}
\usepackage{graphicx}

\def\roughly#1{\mathrel{\raise.3ex
\hbox{$#1$\kern-.75em\lower1ex\hbox{$\sim$}}}}

\newcommand{\be}{\begin{equation}}

\newcommand{\ee}{\end{equation}}
\newcommand{\bqa}{\begin{eqnarray}}
\newcommand{\eqa}{\end{eqnarray}}

\begin{document}
 
\rightline{BI-TP 2019/02}

\vspace{1.0cm}

\title{On gravitational spherical collapse without spacetime singularity}

\author{ R.~Baier} 
\email {baier@physik.uni-bielefeld.de}

\affiliation{Faculty of Physics, University of 
 Bielefeld, D-33501 Bielefeld, Germany}

\vspace{2.5cm}

\begin{abstract}
\vspace{3cm}
This note discusses possible quantum effects in the context of
homogeneous flat FLRW and inhomogeneous LTB
gravitational spherical collapse, which lead to bounce solutions. 
\end{abstract}

\maketitle

\section{Introduction}

In this note the statement: "There is no singularity, no event horizon and no
 information paradox"
given in the context of black holes
by S.~A.~Hayward \cite{Hayward:2005ny},
but also  in \cite{Hawking:2014tga}, is supported by applying effective equations for a flat FLRW \cite{Poisson:2004,dInverno:1992gxs} homogeneous
gravitational collapse, which captures quantum effects of loop quantum gravity (LQG) \cite{Bojowald:2012sp,Rovelli:2018jgs}.
For a discussion of quantum effects see the recent review \cite{Frolov:2014wja}.
According to \cite{Ashtekar:2006wn} the presented modification of the classical equations in the homogeneous case can be interpreted by assuming that Newton's constant is given by
$G_{eff} = G ( 1-\rho/\rho_{crit})$, with the critical density
$\rho_{crit} \propto \rho_{Planck} \propto O(1/G^2 \hbar)$ \cite{Ashtekar:2006wn}.
In more detail the presented analysis is slightly extending and motivated by
the work \cite{Bambi:2013caa,Bambi:2013gva,Malafarina:2018pmv}
e.g. by also including the case of naked singularities
(in the classical limit \cite{Joshi:2008zz}).
Possible bouncing solutions are discussed in the inhomogeneous
LTB \cite{Tolman:1934za,Bondi:1947fta} dust
case \cite{Liu:2014kra},
showing the absence of e.g. naked singularities.

Recent reviews including cosmology are e.g.
\cite{Singh:2015fha,Malafarina:2017csn}.

\section{Quantum collapse}

In order to describe gravitational collapse
\cite{Visser:2009xp,Joshi:2002} a scalar field model coupled to gravity with a scalar field $\Phi(t)$ is assumed.
The resulting energy density $\rho(t)$ and pressure $p(t)$ satisfy the equation of state given by
\be
p = w \rho,~~ (1 + w) = \frac{2}{3} ( 1 - \beta), ~ \beta < 1.
\label{1a}
\ee
The flat homogeneous  FLRW metric \cite{Poisson:2004,dInverno:1992gxs} is used
\be
ds^2 = - dt^2 + a^2(t) [dr^2 + r^2 d\Omega^2] ~,
\label{2a}
\ee
where $a(t)$ is the scale factor.
 Taking into account quantum effects to order $\rho^2$ in the FLRW equation,
 reading \cite{Ashtekar:2006wn},
 \be
 H^2 = (\frac{\dot{a}}{a})^2 = \frac{G}{3}
 \rho(t) ( 1 - \frac{\rho}{\rho_{crit}})~,
 \label{3a}
 \ee
 with $\rho_{crit} \propto O(1/G^2 \hbar)$ and $H(t)$ the Hubble parameter,
 both depending on $\hbar$.
 As a backreaction an ingoing negative energy flux is actually present
 \cite{Mersini-Houghton:2014zka}.
 The limit $\rho_{crit} >> \rho$, i.e. $\hbar \to  0$ implies
 the classical general limit (see for a summary and notation
 \cite{Baier:2014ita,Baier:2015hzl}). The continuity equation reads
 \be
 \dot{\rho}(t) = - 3 H(t) (\rho + p) = - 3 H(t)(1 + w) \rho(t).
 \label{4a}
 \ee
Eqs. \ref{3a} and \ref{4a} lead to the dependent Einstein equation
\be
\dot{H}(t) = - \frac{G}{2} ( \rho + p) ( 1 - \frac{2 \rho}{\rho_{crit}}).
\label{5a}
\ee
Note $\rho_{crit}/3 = 1/{a_{cr}^3}$ in \cite{Bambi:2013caa},
and in the following $\rho_{crit}/3 = \hat{\rho}$.

The analytic solutions for $a(t)$ and $\rho(t)$ (in units $G=c=1$) 
are
\be
a(t) = a_B [(\hat{\rho} -1) ( 1 - t/t_B)^2 + 1]^{\frac{1}{2 (1-\beta)}}~,
\label{6a}
\ee
\be
a_B = \hat{\rho}^{-\frac{1}{2(1-\beta)}}~,
\label{7a}
\ee
such that $a(t=0) = 1$ and $a(t = t_B) = a_B \ne 0$,
with
\be
t_B = \frac{\sqrt(\hat{\rho} - 1)}{(1-\beta)\sqrt{\hat{\rho}}}.
\label{8a}
\ee

\noindent
$a(t)$ does not vanish as a function of $t$, instead it bounces at $t = t_B$
\cite{Bambi:2013caa,Bambi:2013gva}; see also
\cite{Visser:2009xp,Malafarina:2018pmv,Rovelli:2014cta,Cai:2014zga,deHaro:2012cj}.
For $\rho_{crit} \to \infty$, the classical gravity limit
\cite{Joshi:2011hb,Joshi:2008zz,Joshi:2007zza}
is obtained,
\be
a_{cl}(t) = (1 -t/t_s)^{\frac{1}{(1-\beta)}} ~,
\label{9a}
\ee
with $t_s = 1/(1-\beta)$ ~,
such that
\be
t_B = t_s \sqrt{\frac{(\hat{\rho} - 1)}{\hat{\rho}}} < t_s ~.
\label{10a}
\ee
(see Fig.1). Both $a_B$ and $t_B$ depend on $\hbar$.

\begin{figure}[ht]
\psfig{
bbllx=120,bblly=496,bburx=400,bbury=690,
file=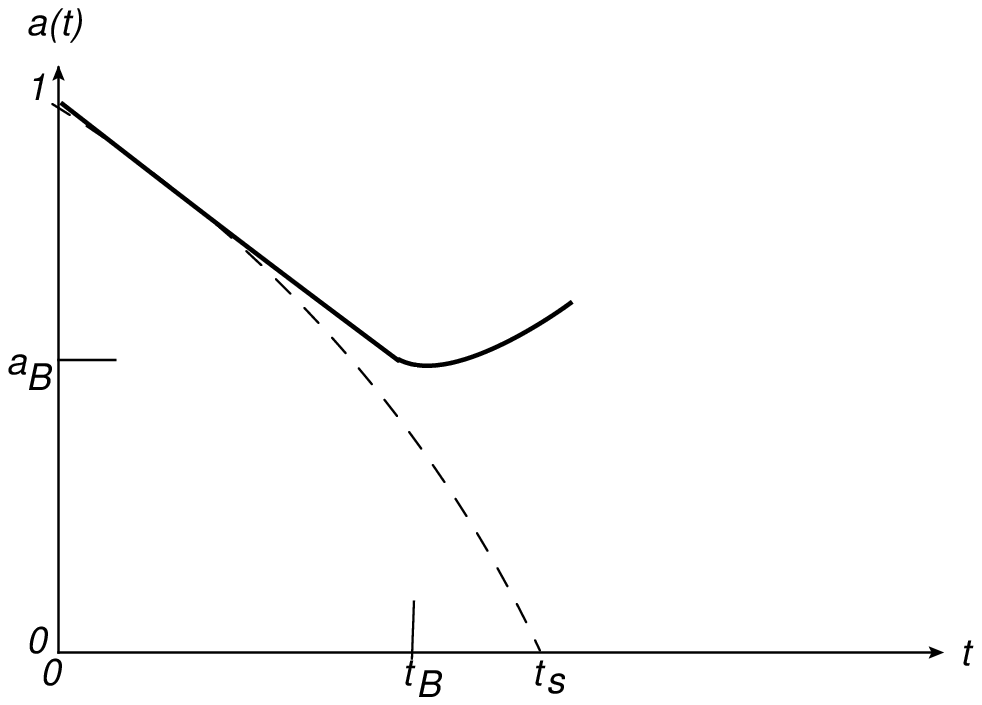,width=8.0cm}
\end{figure}

\vspace{0.8cm}
Fig.1 ~Schematic illustration of the
scale factor: the bouncing  $a(t)$, eq.(\ref{6a}) (solid curve)
        and the classical model $a_{cl}(t)$, eq.(\ref{9a}) (dashed curve).
	See also Fig.1 in
	\cite{Bambi:2013caa} and Fig.2 in \cite{Liu:2014kra}.

\vspace{0.8cm}

\noindent
The density $\rho(t)$ reads with $\rho(0) = 3$,
\be
\rho(t)/3 = \frac{\hat{\rho}}{[(\hat{\rho} - 1)(1 - t/t_B)^2 +1]}~,
\label{11a}
\ee
which is finite at $t = t_B$, namely $\rho(t_B) = \rho_{crit}$.
In the classical limit it becomes
\be
\rho_{cl}(t)/3 = \frac{1}{(1- t/t_s)^2} ~,
\label{12a}
\ee
which is divergent at $t = t_s = \frac{1}{1- \beta}$.

In summary the fluid is first collapsing for $0 \le t < t_B$, and than for $t \ge t_B$ expanding, i.e. a bouncing situation. At $t = t_B$, the theory becomes free according to eq.(\ref{3a}), i.e. $H = 0$.
The Hubble constant reads
\be
H(t) = - \frac{\sqrt{\hat{\rho}(\hat{\rho} - 1)} ( 1 - t/t_B)}{[(\hat{\rho} - 1) ( 1 - t/t_B)^2 + 1]}~.
\label{13a}
\ee

\section{Apparent horizon}

Introducing $R(r,t) = r a(t)$, the physical radius of the bouncing matter, the location of the apparent horizon is given by 
\cite{Joshi:2007zza,Hayward:1994,Hayward:1996,Hayward:2000ca,Booth:2005qc,Ashtekar:2004cn}
\be
R_{ah}^2 = \frac{1}{H^2(t)}~,
\label{14a}
\ee
i.e.
\be
r_{ah}^2 = \frac{1}{{\dot a}^2(t)}~,
\label{15a}
\ee
derived from the expansion
\be
\Theta = \Theta_{+} \Theta_{-} = 2 (H^2(t) - \frac{1}{R^2}) = 0 ~.
\ee
For the collapsing (expanding) phase one requires for the location
\be
\Theta_{+} = 0 ~, i.e.~ R_{ah} = - \frac{1}{H(t)}
\label{17a}
\ee
versus
\be
\Theta_{-} = 0 ~, i.e. ~R_{ah} = + \frac{1}{H(t)}~,
\label{17aa}
\ee
respectively.
Eq.(\ref{14a}) expressed in terms of $t/t_B$
reads
\be
(t/t_B)_{ah} = 1 \mp
\frac{\sqrt{\hat{\rho}}}{2 \sqrt{\hat{\rho} -1}} R_{ah}
\mp \sqrt{( \frac{{\hat{\rho}}}{4 ({\hat{\rho} -1})} R_{ah}^2 -
\frac{1}{{\hat{\rho} -1}})}~,
\label{18a}
\ee
which is except for $t_B$ independent on $\beta$ and
where the first $(+)$ sign corresponds to the case (\ref{17aa}).

From the expression of the vanishing square root in eq.(\ref{18a})
a minimum radius of the horizon in the $(t,R)$ plane is derived
\be
R_{ah}^{min} = \frac{2}{\sqrt{\hat{\rho}}}
\label{19a}
\ee
at
\be
(t/t_B)_{ah}^{min} = 1 \mp \frac{1}{\sqrt{\hat{\rho} -1}}~.
\label{20a}
\ee
The boundary at $t_{ah}^{min}$ is evaluated by
\be
R_b(t_{ah}^{min}) = r_b [\frac{2}{\hat{\rho}}]^{\frac{1}{2(1-\beta)}} ~,
\label{21a}
\ee
with $r_b$ the boundary of the fluid.

For $R_b(t_{ah}^{min}) > R_{ah}^{min}$ there are trapped regions, sketched
in Fig.2,
whereas for $R_b(t_{ah}^{min}) < R_{ah}^{min}$ there are no trapped
regions formed in the physical region of the bouncing fluid. 

In terms of the coordinate $r$
one may introduce a critical $r_c$ by $r_b = r_c$ with
$R_b(t_{ah}^{min}) = R_{ah}^{min}$, which reads 

\be
r_c  = 2^{\frac{1 -2 \beta}{2 ( 1- \beta)}}
{\hat{\rho}}^{\frac{\beta}{2 ( 1 - \beta)}}~.
\label{22a}
\ee
Assuming the physical region of the bouncing fluid (cloud) to be given by
$ r < r_b$, with a small boundary radius $r_b << 1$, then for
\be
r_b < r_c
\label{25a}
\ee
there is no trapped surface formed in contrast with the condition
\be
r_b > r_c~,
\label{26a}
\ee
which allows to form a trapping region (see Fig.2).

\vspace{2cm}

\begin{figure}[ht]
\psfig{
bbllx=75,bblly=475,bburx=430,bbury=720,
file=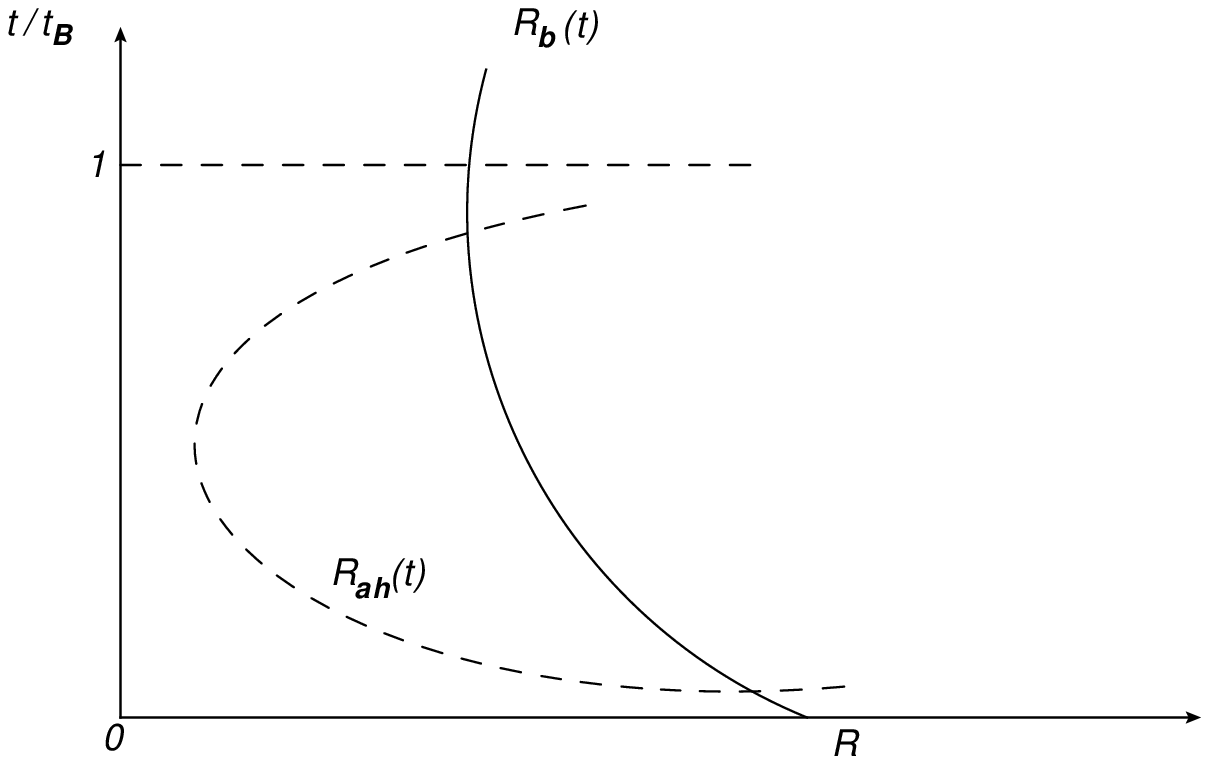,width=9.0cm}
\end{figure}

\vspace{0.8cm}

Fig.2 ~Inner
apparent horizon and trapped region in the collision phase. The expanding phase is obtained by reflecting the curves on the line $t/t_B = 1$.

\vspace{0.8cm}

A numerical example is with $\hat{\rho} = 1000$ \cite{Bambi:2013caa} and for the classical black hole cases:

- dust \cite{Oppenheimer:1939ue} with $\beta = -1/2$
\be
r_c =  0.5~,
\ee
i.e. no trapped region for $r_b \le 0.5$.

For the classical naked singularity case, e.g. for $\beta = + 1/2$,
\be
r_c  = \sqrt{\hat{\rho}} >> 1~,
\ee
which is much larger than $r_b < 1$, and no trapped region inside the
fluid is present.

The case of the presence of naked singularities in the classical limit
\cite{Joshi:1993zg,Joshi:2001xi,Joshi:2008zz} is changed, no singularity is present for $\beta > 0$ in the interior region after the critical density $\rho_{crit} (\hat{\rho})$ is introduced.

\section{Exterior metric}

In order to obtain the full spacetime (here for $\beta < 0$),
the metric eq.(\ref{2a}) has to be matched with the outer
region $r \ge r_b$.

Convenient coordinates are the advanced Eddington-Finkelstein ones
\cite{Poisson:2004}.
In the following only the collapsing sector $t/t_B \le 1$ is discussed; the expanding one is obtained by symmetry with respect to the line $t/t_B = 1$.

Introducing the transformation with $R = r a(t)$,
\be
a dr = \frac{1}{1 - RH} dR - RH dv ~,
\label{1c}
\ee
\be
dt = - \frac{1}{1 - RH} dR + dv ~,
\label{2c}
\ee
leading to
\be
ds^2 = - (1 - R^2 H^2) dv^2 + 2 dv dR +R^2 d\Omega^2~.
\label{3c}
\ee
In order to perform the matching at the boundary $r = r_b$
it is usefull to sketch it by using analytical approximations.
In the following a convenient one is obtained by the limit of
large negative $\beta$ ($\beta \rightarrow - \infty$),
i.e. $ a(t) \rightarrow 1$, but $H(t)\approx \dot{a}(t) \ne 0$
(eq.(\ref{13a})), and $t_B \ne 0$. It implies $R_B(t) \approx r_b$.

\noindent
The points $(+)$ and $(-)$ in Fig.3 are determined by eq.(\ref{18a})
with $R_{ah} = R_b \approx r_b$,
\be
(t/t_B)_{\pm} \approx 1 -
\frac{\sqrt{(\hat{\rho}} r_b)}{2 \sqrt{(\hat{\rho} - 1)}}
\pm \frac{\sqrt{\hat{\rho}} r_b^2 - 4}{2 \sqrt{(\hat{\rho} - 1)}}~.
\label{4c}
\ee

\begin{figure}[ht]
\psfig{
bbllx=95,bblly=450,bburx=435,bbury=670,
file=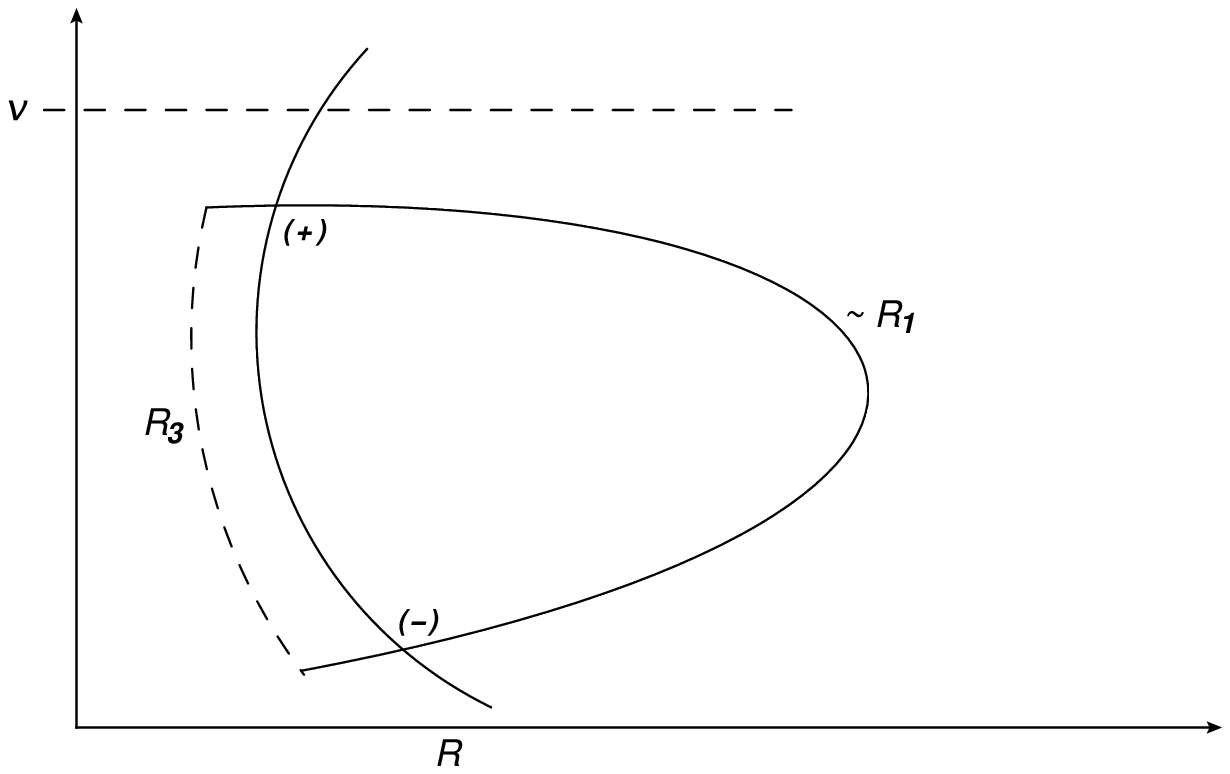,width=9.0cm}
\end{figure}

\vspace{0.8cm}

Fig.3 Outer apparent horizon and trapped region (see Fig.2
in \cite{Binetruy:2018jfz}).

\newpage

\noindent
For the exterior metric the coordinates
\be
ds^2 = - F(v,R) dv^2 + 2 dv dR +R^2 d\Omega^2~
\label{5c}
\ee
are introduced, and the Hayward metric
\cite{Hayward:2005gi,Binetruy:2018jfz}
is used for illustration
\be
F(v,R) = 1 - \frac{2 m(v) R^2}{R^3 + 2m(v) l^2}~~, ~ l = cst.~
\label{6c}
\ee






\noindent
Following \cite{Binetruy:2018jfz} for $m(v)$ the form
\be
2 m(v) = cst \exp{(\frac{ (v - v_0)^2}{\sigma^2})}
\label{7c}
\ee
is taken. The location of the horizons (see Fig.3)
is obtained from $\Theta_+ = 0$,
i.e.
$F(v,R)=0$, or
\be
R^3 - 2 m(v) R^2 + 2m(v) l^2 = 0~,
\label{8c}
\ee
with the approximate solutions \cite{Binetruy:2018jfz}
\be
R_1 \approx 2 m + O(l^2/2m)~, R_3 \approx l + O(l^2/2m)~.
\label{9c}
\ee
Matching at the boundary may be obtained by $R_3 \approx l < r_b$ and
$R_1 \approx r_b$ (see Fig.3), i.e.
\be
v_{\pm} \approx v_0 \pm \sigma \sqrt{\ln(\frac{cst}{r_b})}~.
\label{10c}
\ee
On the boundary the $(\pm)$ points are related by
solving
\be
(\frac{dv}{dt})_b \approx \frac{1}{1 - r_b H(t)}~,
\label{11c}
\ee
which finally fixes the parameters $l, cst, v_0$ in terms
of $\hat{\rho}, r_b$.
For illustration one may consider the case of large $\hat{\rho}$,
with $v \approx t$, such that
\be
\sigma \sqrt{\ln(\frac{cst}{r_b})} \approx \frac{r_b t_B}{2} ~,
v_0 \approx (1 - r_b/2) t_B~,
\label{12c}
\ee
with $t_B$ fixed.

\newpage

\begin{figure}[ht]
\psfig{
bbllx=90,bblly=440,bburx=375,bbury=690,
file=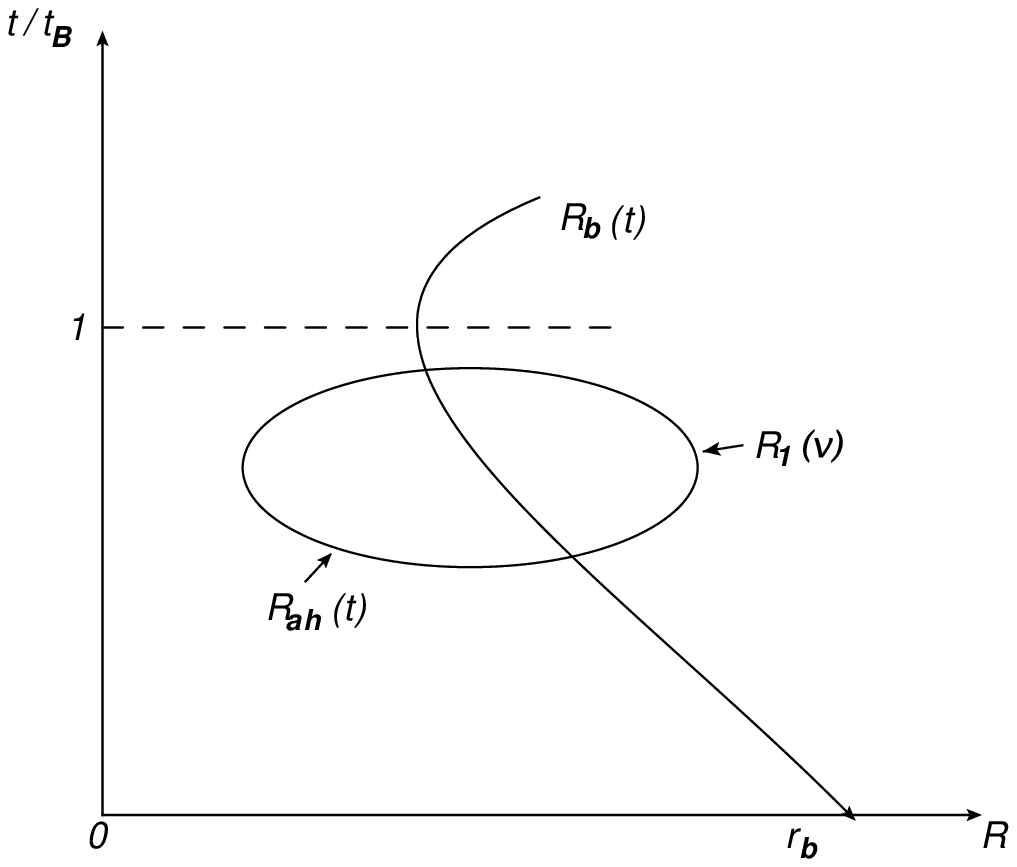,width=9.0cm}
\end{figure}

\vspace{0.8cm}

Fig.4~ Collapse region: sketch for matching the inner and outer regions
with respect to the boundary $R_b(t)$
(see text for details).

\vspace{0.8cm}

This  Fig.4 showing matching indicates that only closed trapping
horizons exist for a finite time in the discussed non-singular
spacetime geometry.

\section{Inhomogeneous dust}

By describing inhomogeneous dust (pressure $p = 0$, $\beta = - 1/2$,
extending \cite{Oppenheimer:1939ue})
the Lema{\^i}tre-Tolman-Bondi (LTB) flat metric \cite{Tolman:1934za,Bondi:1947fta}
\be
ds^2 = - dt^2 + R'^2 dr^2 + R^2 d\Omega^2
\label{1i}
\ee
is used with $ R = R(t,r)$ and $R' = \frac{\partial}{\partial r} R(t,r)$.
The classical Einstein field equation reads
\be
\frac{{\dot{R}}^2}{R^2} = \frac{{\dot a^2 (t,r)}}{a^2 (t,r)} = \frac{F(r)}{R^3}~,
\label{2i}
\ee
where the mass $F(r) = r^3 M(r)$ and $R(t,r) = r a(t,r)$ are introduced
\cite{Joshi:2014gea}.

\noindent
In comparison of this equation  with eq.(\ref{3a}) the average density is defined by
\be
\rho_{av}(t,r) = \frac{3 F(r)}{R^3} = \frac{3 M(r)}{a^3(t,r)}, ~G =1~.
\label{3i}
\ee
Possible quantum effects to order $\rho_{av}^2$ are introduced in analogy to eq.(\ref{3a}) by modifying eq.(\ref{2i}), introducing an effective density
$\rho_{effective}$ \cite{Malafarina:2017csn},
\be
\frac{{\dot{R}}^2}{R^2} =
\frac{1}{3}~ \rho_{effective} (t,r) =\frac{1}{3} \rho_{av}
( 1 - \frac{\rho_{av}}{\rho_{crit}}) =
\label{4i}
\ee
\be
= \frac{M(r)}{a^3 (t,r)} ( 1 - \frac{M(r)}{a^3 (t,r) \hat{\rho}})~,
\hat{\rho} = \rho_{crit}/3 = const.
\label{5i}
\ee
This equation is solved by a non-vanishing scale factor
\be
a(t,r) = \{ \frac{M(r) + (\hat{\rho} - M(r)) (1 - t/t_B)^2)}{\hat{\rho}}\}^\frac{1}{3}~,
\label{6i}
\ee
with $a(t=0,r) = 1$ and which bounces at
\be
a_B (t_B,r) = (\frac{M(r)}{\hat{\rho}})^{\frac{1}{3}} \ne 0
\label{7i}
\ee
at
\be
t_B(r) = \frac{2}{3} \sqrt{\frac{\hat{\rho} - M(r)}{\hat{\rho} M(r)}}~,
\label{8i}
\ee
where $t_B \ge 0, \hat{\rho} \ge M(r)$.

Assuming $M(r) > 0$, but $M'(r) <0$, e.g. $M(r) \simeq 1 - M_2 r^2 > 0$,
i.e. a decreasing mass with increasing radius $r$,
the bounce time $t_B (r)$ $(a_B (t_B, r))$ is increasing (decreasing) with $r$,
near $r \simeq 0$ 
(see Fig.5).

The crucial result is a finite density \cite{Joshi:2014gea}
\be
\rho(t,r) = \frac{F'}{R^2 R'}~,
\label{9i}
\ee
which behaves near $r \simeq 0$ as
\be
\rho(t, r \simeq 0) \simeq \frac{3 M(r)}{a^3(t,r)}\vert_{r \simeq 0}
   = \rho_{av}(t, r \simeq 0) =
\ee
\be
 = \frac{3 \hat{\rho}}{ 1 + (\hat{\rho} -1) ( 1 - t/t_B)^2}~,
\label{10i}
\ee
with $M(r \simeq 0) \simeq 1$.
$\rho$ is finite at $t_B$, but becomes, however, singular at $t = t_s$ in the classical limit
$\hat{\rho} \rightarrow \infty$.

The case of section I. is reproduced with $M(r) = 1$ and $\beta = -1/2$.
In both cases the scale factor as a function of $t$ does not vanish, instead it bounces at $ a = a_B$ at $t_B$, where in the inhomogeneous dust case 
both functions depend on $r$.

In the classical limit, $\hat{\rho} \rightarrow \infty$, the radius $R_{cl} (t,r)$
reads \cite{Joshi:2014gea}
\be
R_{cl}(t,r) = r ( 1 - t/t_s(r))^{\frac{2}{3}} =
   [\frac{3}{2} \sqrt{F(r)}(t_s(r) - t)]^{\frac{2}{3}}~,
\label{11i}
\ee
which vanishes at $t_s(r) = \frac{2}{3 \sqrt{M(r)}}$. 
Because of the $r$ dependence there is no simultaneous collapse.

As an example of a collapse scenario: the dust has a boundary at $r = r_b > 0$,
and $M(0) > M(r_b)$, i.e. a maximum mass at $r = 0$, with $M'(r) < 0$, $t_s' = - \frac{M'}{3 M^{3/2}} > 0$. See e.g. Fig.2 in \cite{Malafarina:2011pe}.
There is a formation of a locally naked singulariy
\cite{Joshi:1993zg,Joshi:2014gea}. For a review \cite{Joshi:2007zza}.

\noindent
The homogeneous singular limit is finally obtained with $M(r) = 1,t_s = 2/3$
(see eq.(\ref{9a})).

To generalize section II. for the inhomogeneous case the apparent horizon is defined by $g^{\mu \nu} \partial_\mu R \partial_\nu R = 0$, i.e.
\be
{\dot{R}}^2_{ah} = \frac{F}{R} (1 - \frac{F}{R^3 \hat{\rho}}) = 1~,
\label{12i}
\ee
i.e. collapsing phase: $\dot{R}_{ah} = -1$ and expanding phase: $\dot{R}_{ah} = + 1$.
In the following the collapsing phase up to the bouncing time $t_B$
is considered,
\be
{\dot{R}}_{ah} = - \frac{r \sqrt{M(r) (\hat{\rho} - M(r))}(1 - t/t_B)}{a^2(t,r)
\sqrt{\hat{\rho}}} = - 1~.
\label{13i}
\ee
Note that ${\dot{R}}_{ah}(t_B,r) = 0$.

Eq.(\ref{13i}) can be written  for $t = t_{ah}$ and $r_{ah}$ as
\be
(1 - t/t_B)^2 - (r_{ah} \sqrt{M})^{3/2} (\frac{\hat{\rho}}{\hat{\rho} - M})^{1/4}
(1 - t/t_B)^{3/2} + \frac{M(r)}{\hat{\rho} - M} = 0~.
\label{14i}
\ee
In the classical limit $\hat{\rho} \rightarrow \infty$ the apparent horizon
is given by \cite{Joshi:2014gea}
\be
t_{ah}(r) = t_s(r) - \frac{2}{3} r^3 M(r)~,
\label{15i}
\ee
in the flat region of the marginally bound collapse.
It is noted \cite{Joshi:2014gea} that the collapse is simultaneous
for $t_s(r) = cst.$, i.e. $M=1 = cst.$,
$t_{ah} = \frac{2}{3} ( 1 - r ^3)$, which corresponds to the
Oppenheimer-Snyder collapse \cite{Oppenheimer:1939ue}.

Having introduced a finite critical density $\rho_{crit}$ as a possible result due to quantum effects the scale factor $a(t,r)$, eq.(\ref{6i}), doesnot vanish
 $a(t,r) \ne 0$ for $M(r) \ne 0$, and the classical singular collapse, a black hole or a naked singularity, is replaced by a system which bounces back
 ($\dot{a}(t,r) = 0$ at $t_B$) before it could reach the singularity (Fig.5).

\vspace{0.8cm}

\begin{figure}[ht]
\psfig{
bbllx=130,bblly=430,bburx=495,bbury=660,
file=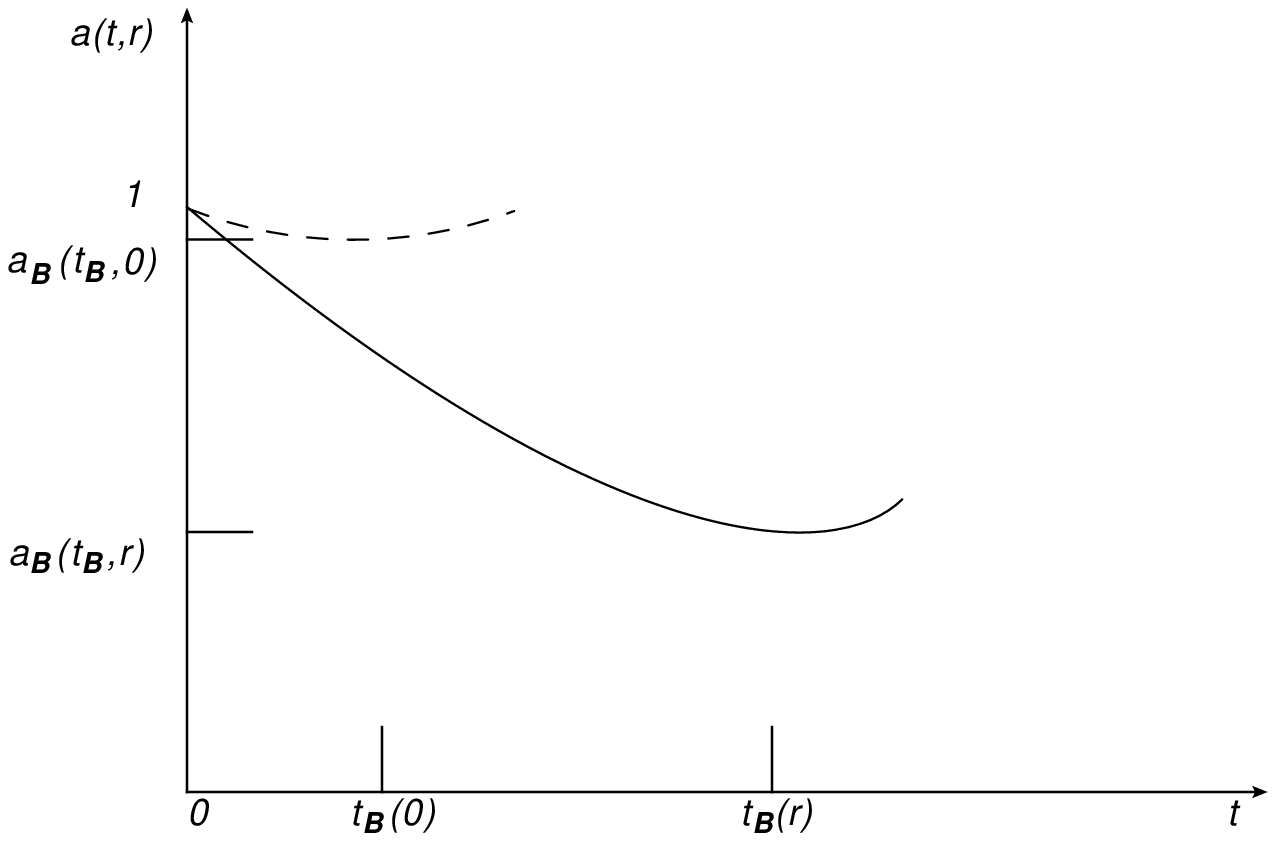,width=9.0cm}
\end{figure}

\vspace{0.8cm}

Fig.5~ Schematic  illustration of the non-vanishing
scale factor  $a(t,r) $ of eq.(\ref{6i}) as a function of $t$.
Solid (dashed) curve for $t > 0$ $(t = 0)$.
\vspace{0.8cm}

There is even no trapped region present inside the cloud bounded by $r = r_b$,
when the condition is satisfied, which generalizes the homogeneous  condition
in the $r,t$ plane given in eq.(\ref{25a})
for $\beta = -1/2$ (dust),
\be
 r_{ah}^{min} > r_b~.
\label{17i}
\ee
To indicate in an approximate way for small values of $r_b$ and $r$,
and  $M/\rho \simeq M(r_b)/\rho = const$ and small, it reads
from eq.(\ref{14i}),
 \be
 r_{ah}^{min} \simeq \frac{2^{4/3}}{\hat{\rho}^{1/6} \sqrt(3 M(r_b))} 
    ~.
 \label{18i}
 \ee
As long as $M(r_b) < 1$ a trapped inside region becomes more unlikely in the
inhomogeneous than in the homogeneous case.

\vspace{1cm}

\centerline{\bf{Acknowledgements}}

\vspace{0.3cm}

\noindent
Special thanks to S. von Reder for help with drawing the figures.

\vspace{1cm}


\end{document}